\documentclass[11pt,pdftex, preprint]{elsarticle}
\usepackage[usenames,dvipsnames,svgnames,table]{xcolor}
\usepackage[colorlinks]{hyperref} 
\AtBeginDocument{\hypersetup{linkcolor=blue, citecolor=blue}} 
\usepackage[T1]{fontenc}
\usepackage[font=footnotesize,labelfont=bf]{caption}
\usepackage{bm, bbm, bbding} 
\usepackage{seqsplit}
\usepackage{amssymb,amsmath,amsfonts,mathrsfs,graphicx,caption,kantlipsum,units,yfonts}
\usepackage{txfonts,dsfont}
\usepackage{feynmf} 
\usepackage{pdflscape}
\usepackage{rotating} 
\usepackage[toc,page,title,titletoc,header]{appendix}
\usepackage{lineno,setspace, framed, hyperref}
\usepackage[top=2.6cm, bottom=2.7cm, left=2.8cm, right=2.8cm]{geometry}
\usepackage[novbox]{pdfsync} 
\usepackage{subcaption}

\usepackage[numbers]{natbib}

\makeatletter
\def\ps@pprintTitle{%
  \let\@oddhead\@empty
  \let\@evenhead\@empty
  \def\@oddfoot{\reset@font\hfil\thepage\hfil}
  \let\@evenfoot\@oddfoot
}
\makeatother

\journal{}
\begin{document}

\begin{frontmatter}

\title{Thermal relics as hot, warm and cold dark matter in power-law $f(R)$ gravity}
\author{David Wenjie Tian\fnref{myfootnote}}
\address{Faculty of Science,  Memorial University, St. John's, Newfoundland, A1C 5S7, Canada}
\fntext[myfootnote]{Email address: wtian@mun.ca}

\begin{abstract}
We investigate the thermal relics as hot, warm and cold dark matter in 
$\mathscr{L}=\varepsilon^{2-2\beta}R^\beta+{16\pi}m_{\text{Pl}}^{-2}\mathscr{L}_m$ gravity, where 
$\varepsilon$ is a constant balancing the 
dimension of the field equation, and 
$1<\beta<(4+\sqrt{6})/5$ for the positivity of energy 
density and temperature. If light neutrinos serve as hot/warm relics, the entropic number of statistical degrees of freedom $g_{*s}$ at freeze-out and thus the predicted fractional energy density $\Omega_\psiup h^2$ are $\beta-$dependent, which relaxes the standard mass bound $\Sigma m_\nu$. For cold relics, by exactly solve the simplified Boltzmann equation in both relativistic and nonrelativistic regimes, we show that the Lee-Weinberg bound for the mass of heavy neutrinos can be considerably relaxed, and the ``WIMP miracle'' for weakly interacting massive particles (WIMPs) gradually invalidates as $\beta$ deviates from $\beta=1^+$.    The whole framework reduces to become that of GR in the limit $\beta\to 1^+$.\\

\end{abstract}

\end{frontmatter}


\section{Introduction}

With the development of observational astrophysics and cosmology, the investigations of galaxy rotation curves, gravitational lensing and large scale structures have provided strong evidences for the existence and importance of dark matter. The abundance of dark matter
has been measured with increasingly high precision, such as $\Omega_{\text{dm}} h^2=0.1198\pm0.0026$ by the latest Planck data \cite{Planck data}; however, since our knowledge of dark matter exclusively comes from the gravitational effects, the physical nature of dark-matter particles remain mysterious.

Nowadays it becomes a common view that to account for the observed dark
matter, 
one needs to go beyond the SU$(3)_c\times$SU$(2)_W\times$U$(1)_Y$ minimal standard model. There are mainly two leading classes of dark-matter candidates: axions that are non-thermally 
produced via quantum phase transitions in the early universe, and generic weakly interacting massive particles (WIMPs) \cite{Lee Weinberg bound} 
that freeze out of thermal equilibrium from the very early cosmic plasma and leave a relic density matching the
present-day Universe. In this paper, we are interested in the latter class, i.e. dark matter created as thermal relics. We aim to correct and complete the pioneering investigations in Ref.\cite{BBN fR II} 
for cold relics in $\mathscr{L}=m_{\text{Pl}}^{2-2\beta}R^\beta+{16\pi}m_{\text{Pl}}^{-2}\mathscr{L}_m$ gravity, and provide a comprehensive investigation of thermal relics as hot, warm and cold dark matter in $\mathscr{L}=\varepsilon^{2-2\beta}R^\beta+{16\pi}m_{\text{Pl}}^{-2}\mathscr{L}_m$  gravity.


This paper is organized as follows. Sec.~\ref{Gravitational framework of power-law fR gravity} sets up the gravitational framework of $\mathscr{L}=\varepsilon^{2-2\beta}R^\beta+{16\pi}m_{\text{Pl}}^{-2}\mathscr{L}_m$ gravity, while Sec.~\ref{Preparations Thermal relics} generalizes the time-temperature relation for cosmic expansion and derives the simplified Boltzmann equation. Sec.~\ref{Hot warm relic dark matter and light neutrinos} studies hot/warm thermal relics, and shows the influences of $\beta$ and $\varepsilon$ to the bound of light neutrino mass. Sec.~\ref{Cold relic dark matter} investigates cold thermal relics by solving the simplified Boltzmann equation, while Sec.~\ref{Example Fourth generation massive neutrinos and Lee-Weinberg bound} rederives the Lee-Weinberg bound on fourth-generation massive neutrinos, and examines the departure from electroweak energy scale. Finally, the GR limit of the whole theory is studied in Sec. 7.

Throughout this paper, for the physical quantities involved in the calculations of thermal relics, we use the 
natural unit system of particle physics which sets $c=\hbar=k_B=1$ and is related to le syst\`eme 
international d'unit\'es by $1\text{ MeV}=1.16\times 10^{10}\text{ kelvin} = 
1.78\times 10^{-30}\text{ kg}=(1.97\times 10^{-13} \text{ meters})^{-1}= (6.58\times 10^{-22}\text{ seconds})^{-1}$. On the other hand, for the spacetime geometry, we adopt the conventions 
$\Gamma^\alpha_{\beta\gamma}=\Gamma^\alpha_{\;\;\,\beta\gamma}$,
$R^{\alpha}_{\;\;\beta\gamma\delta}=\partial_\gamma \Gamma^\alpha_{\delta\beta}\cdots$ and 
$R_{\mu\nu}=R^\alpha_{\;\;\mu\alpha\nu}$ with the metric signature $(-,+++)$.


\section{Gravitational framework of power-law $f(R)$ gravity}\label{Gravitational framework of power-law fR gravity}

$f(R)$ gravity is a direct generalization of GR and extends the 
Hilbert-Einstein action $I_{\text{HE}}=\int \!\sqrt{-g}\,d^4x$ $ \left(R+{16\pi}m_{\text{Pl}}^{-2}\mathscr{L}_m\right)$ into
\begin{equation}
\label{f(R) action}
\mathcal{I}=\int d^4x\!\sqrt{-g}\,\left[f(R,\varepsilon)
+{16\pi}m_{\text{Pl}}^{-2} \mathscr{L}_m\right]\,,
\end{equation}
where $R$ denotes the Ricci scalar of the spacetime, $\varepsilon$ is some constant 
balancing the dimensions of the field equation, and $\mathscr{L}_m$ is the matter Lagrangian density.
Also, $m_{\text{Pl}}$ refers to the Planck mass, 
which is related to Newton's constant $G$ by
$m_{\text{Pl}}\coloneqq 1/\!\sqrt{G}$ and takes the value $m_{\text{Pl}}\simeq 1.2209\times 10^{22}\text{ MeV}$.

Variation of Eq.(\ref{f(R) action})
with respect to the inverse metric $\delta \mathcal{I}/\delta g^{\mu\nu}=0$ 
yields the field equation
\begin{equation}
\label{Field Eq generic fR}
f_R R_{\mu\nu}-\frac{1}{2}f+\left(g_{\mu\nu}\Box-\nabla_\mu\nabla_\nu\right)f_R
={8\pi}m_{\text{Pl}}^{-2} \mathcal{T}_{\mu\nu}^{(m)}\,,
\end{equation}
where $f_R\coloneqq df(R,\varepsilon)/dR$, $\Box$ 
denotes the covariant d'Alembertian $\Box\coloneqq g^{\alpha\beta}\nabla_\alpha\nabla_\beta$,  
and $\mathcal{T}_{\mu\nu}^{(m)}\coloneqq \frac{-2}{\!\sqrt{-g}}\frac{\delta\left(\sqrt{-g}\mathscr{L}_m\right)}{\delta g^{\mu\nu}}$ 
is the stress-energy-momentum tensor of the physical matter content. This 
paper considers the spatially flat, homogeneous and isotropic Universe, which, in the $(t,r,\theta,\varphi)$ comoving coordinates along the cosmic Hubble flow, is depicted by the Friedmann-Robertson-Walker (FRW) line element
\begin{equation}
\label{FRW metric}
\begin{split}
ds^2 = -dt^2+ a(t)^2 dr^2 + a(t)^2 r^2
 \Big( d\theta^2+\sin^2 \!\theta d\varphi^2 \Big)\,,
\end{split}
\end{equation}
where $a(t)$ denotes the cosmic scale factor. Assume a perfect-fluid 
material content $\mathcal{T}^{\mu(m)}_{\;\;\nu} = \text{diag} [-\rho, P, P,$ $P]$,
with $\rho$ and $P$ being the energy density and pressure, respectively. Then
Eq.(\ref{Field Eq generic fR}) under the flat FRW metric yields
the generalized Friedmann equations
\begin{equation}
\label{Friedmann EqI generic FR}
3\frac{\ddot{a}}{a}f_R-\frac{1}{2}f
-3\frac{\dot{a}}{a}f_{RR}\dot{R}=-{8\pi}m_{\text{Pl}}^{-2} \rho\,,
\end{equation}
\begin{equation}
\label{Friedmann EqII generic FR}
\left(\frac{\ddot{a}}{a}+2\frac{\dot{a}^2}{a^2} \right)f_R-\frac{1}{2}f- 
f_{RR}\ddot{R}
- f_{RRR}(\dot{R})^2-3\frac{\dot{a}}{a}f_{RR}\dot{R}= {8\pi}m_{\text{Pl}}^{-2} P\,,
\end{equation}
where overdot denotes the derivative with respect to the
comoving time, $f_{RR}\coloneqq d^2f(R,\varepsilon)/dR^2$,
and $f_{RRR}\coloneqq d^3f(R,\varepsilon)/dR^3$. In addition,
the equation of local energy-momentum conservation gives rise to the continuity equation,
\begin{equation}
\label{Continuity Eq}
\nabla^\mu \mathcal{T}_{\mu\nu}^{(m)}=0\quad\Rightarrow\quad
\dot{\rho}+3\frac{\dot{a}}{a}
(\rho+P)=0\,;
\end{equation}
for the very early Universe that is radiation-dominated,
integration of Eq.(\ref{Continuity Eq}) with the equation of state $\rho=3P$
yields that the radiation density is related to the cosmic scale factor by
\begin{equation}
\rho\,=\,\rho_0 a^{-4}\,
\propto\,a^{-4}\,.
\end{equation}

In this paper, we will work with the specific
power-law $f(R)$ gravity 
\begin{equation}
\label{Power law fR action}
\mathcal{I}
=\int d^4x\!\sqrt{-g}\,\left(\varepsilon^{2-2\beta}
R^\beta+{16\pi}m_{\text{Pl}}^{-2}\mathscr{L}_m\right)\,,
\end{equation}%
where $\beta=\text{constant}>0$.
With $\rho=\rho_0 a^{-4}$ 
and $f(R)=\varepsilon^{2-2\beta}R^\beta$,
the generalized first Friedmann equation
(\ref{Friedmann EqI generic FR}) yields 
\begin{equation}
\label{Exact solution}
a=a_0 t^{\beta/2}\propto t^{\beta/2}\quad,\quad H\coloneqq\frac{\dot a}{a}=\frac{\beta}{2t}\,,
\end{equation}
\begin{flalign}
\label{Concrete Friedmann EqI generic FR}
\text{and}\hspace{3.5cm}
\left[\frac{12(\beta-1)}{\beta}H^2\right]^\beta
\frac{\big(-5\beta^2+8\beta-2\big)}{\beta-1}=32\pi
\varepsilon^{2\beta-2}m_{\text{Pl}}^{-2}\rho\,,&&
\end{flalign}
where $H$ refers to the cosmic
Hubble parameter.
Moreover, the weak, strong and
dominant energy conditions for
classical matter fields require the
energy density $\rho$ to be positive definite, and as a consequence, the
positivity of the left hand side of Eq.(\ref{Concrete Friedmann EqI generic FR}) limits $\beta$ to the domain
\begin{equation}\label{beta domain I}
1<\beta<\frac{4+\sqrt{6}}{5}
\lesssim 1.2899\,.
\end{equation}
Note that the Ricci scalar for the flat
FRW metric with $a=a_0 t^{\beta/2}$ reads
\begin{equation}
R=6\left(\frac{\ddot{a}}{a}+\frac{\dot{a}^2}{a^2}\right)
=\frac{3\beta(\beta-1)}{t^2} \,,
\end{equation}
so $R>0$ and $R^\beta$ is always well
defined in this domain.
Moreover, we will utilize two choices of
$\varepsilon$ to balance the dimensions in
$\mathscr{L}=\varepsilon^{2-2\beta}R^\beta+{16\pi}m_{\text{Pl}}^{-2}\mathscr{L}_m$ gravity:
\begin{enumerate}
  \item[(i)]  $\varepsilon=1$ [sec$^{-1}$]. This choice can best respect
and preserve existent investigations in mathematical
relativity for the $f(R)$ class of modified gravity, which have been analyzed
for $\widetilde{\mathscr{L}}=f(R)+16\pi m_{\text{Pl}}^{-2}\mathscr{L}_m$
without caring the physical dimensions.

  \item[(ii)]  $\varepsilon=m_{\text{Pl}} \simeq 0.1854\times 10^{44}$ [1/s],
or $1/\ell_{\text{Pl}}$ where $\ell_{\text{Pl}}=\sqrt{G}$
refers to Planck length. The advantage of this
choice is there is no need to employ extra
parameters outside the mathematical expression
$\widetilde{\mathscr{L}}=f(R)+16\pi m_{\text{Pl}}^{-2}\mathscr{L}_m$.

\end{enumerate}


\section{Thermal relics}\label{Preparations Thermal relics}

\subsection{Time-temperature relation of cosmic expansion}
\vspace{2mm}

For the very early Universe, the radiation energy density $\rho$ attributes
to all relativistic species, which are exponentially greater than those of
the nonrelativistic particles, and
therefore $\rho=\sum \rho_i(\text{boson})+\frac{7}{8}\sum \rho_j(\text{fermion})
=\sum \frac{\pi^2}{30}g_i^{(b)} T_i^4(\text{boson})+\frac{7}{8}\sum
\frac{\pi^2}{30}g_j^{(f)} T_j^4(\text{fermion})$, where $\{g_i^{(b)}, g_j^{(f)}\}$
are the numbers of statistical degrees of freedom for relativistic bosons
and fermions, respectively. More concisely,
normalizing the temperatures
of all relativistic species with respect to photons'
temperature $T_\gamma \equiv T$, one has the generalized
Stefan-Boltzmann law
\begin{equation}\label{rho and g for radiation}
\rho=\frac{\pi^2}{30}g_*T^4\qquad\text{with}\qquad
g_*\coloneqq\sum_{\text{boson}} g_i^{(b)}
\left(\frac{T_{i}}{T}\right)^4
+\frac{7}{8}\sum_\text{fermion} g_j^{(f)}
\left(\frac{T_{j}}{T}\right)^4\,,
\end{equation}
where, in thermodynamic equilibrium, $T$ is the common
temperature of all relativistic particles.
To facilitate the discussion
of thermal relics, introduce a dimensionless variable
\begin{equation}
x\coloneqq \frac{m_\psiup}{T}
\end{equation}
to relabel the time scale, where $m_\psiup$ denotes the mass of dark-matter particles. $x$ is a well defined variable since the temperature monotonically decreases after the Big Bang: Reheatings due to pair annihilations at $T\gtrsim 0.5486 \text{ MeV}=m(e^{\pm})$ only slow down the decrement of $T$ rather than increase $T$ \cite{Reheating}.

Substitute Eq.(\ref{rho and g for radiation}) into Eq.(\ref{Concrete Friedmann EqI generic FR}), and it follows that the cosmic expansion rate is
related to the radiation temperature by
\begin{equation}
\label{Hubble evolution}
\begin{split}
H
&= \sqrt{\frac{\beta}{12(\beta-1)}}\,
\left(\sqrt{\frac{(\beta-1)\,g_*}{-5\beta^2+8\beta-2}}\right)^{1/\beta}
\left(\sqrt{\frac{32\pi^3}{30}}
\frac{T^2}{m_{\text{Pl}}}\right)^{1/\beta}
\varepsilon^{1-\frac{1}{\beta}}\\
&=\sqrt{\frac{\beta}{12(\beta-1)}}\,
\left(\sqrt{\frac{(\beta-1)\,g_*}{-5\beta^2+8\beta-2}}\right)^{1/\beta}
\left(\sqrt{\frac{32\pi^3}{30}}
\frac{m^2}{m_{\text{Pl}}}\right)^{1/\beta}
\varepsilon^{1-\frac{1}{\beta}} x^{-2/\beta}\,,
\end{split}
\end{equation}
which can be compactified into
\begin{equation}
H=H(m)x^{-2/\beta}\quad
\text{with}\quad
H(m)\coloneqq H(T=m_\psiup)\,.
\end{equation}
As time elapses after the Big Bang,
the space expands
and the Universe cools. Eq.(\ref{Hubble evolution}) along with $H=\beta/(2t)$
leads to $t=\frac{\beta}{2H}= \frac{\beta x^{2/\beta}}{2H(m)}$ and the time-temperature relation
\begin{equation}\label{WIMP time-temperature relation}
\begin{split}
t&=\sqrt{3\beta(\beta-1 )}\, \left(\sqrt{\frac{ -5\beta^2+8\beta-2 }
{ (\beta-1)\,g_*}}\right)^{1/\beta}
\left(\sqrt{\frac{30}{32\pi^3}}
\frac{m_{\text{Pl}}}{T^2}\right)^{1/\beta} \varepsilon^{1/\beta -1}\\
&=\sqrt{3\beta(\beta-1 )}\, \left(\sqrt{\frac{ -5\beta^2+8\beta-2 }
{ (\beta-1)\,g_*}}\right)^{1/\beta}
\left(\sqrt{\frac{30}{32\pi^3}}\frac{m_{\text{Pl}}}{m_\psiup^2}\right)^{1/\beta} \varepsilon^{1/\beta -1} x^{2/\beta}\,.
\end{split}
\end{equation}

\subsection{Boltzmann equation}
\vspace{2mm}

For dark-matter particles $\psiup$ in the very early Universe (typically before the era of primordial nucleosynthesis), there are
various types of interactions determining the $\psiup$ thermal relics, such as elastic scattering between
$\psiup$ and standard-model particles, and  self-annihilation
$\psiup+\psiup\rightleftharpoons \psiup+\psiup+\cdots$. In this paper, we are interested in $\psiup$ initially in thermal equilibrium via the pair annihilation into (and creation from) standard-model particles $\ell=\gamma, e^{\pm}, \mu^\pm, \tau^\pm\cdots$,
\begin{equation}\label{WIMP pair annihilation creation}
\psiup+\bar\psiup\rightleftharpoons \ell+\bar\ell\,.
\end{equation}
As the mean free path of $\psiup$ increases along the
cosmic expansion, the interaction rate $\Gamma_{\psiup\bar\psiup}$ of Eq.(\ref{WIMP pair annihilation creation})
gradually falls below the Hubble expansion rate $H$, and the abundance of $\psiup$ freezes out.
The number density of $\psiup$ satisfies the simplified Boltzmann equation
\begin{equation}\label{simplified Boltzmann equation}
\dot{n}_\psiup+3Hn_\psiup=-\left\langle\sigma v \right\rangle\left[n_\psiup^2-\left(n_\psiup^{\text{eq}}\right)^2\right]\,,
\end{equation}
where $\left\langle\sigma v \right\rangle$
is the thermally averaged cross-section.
Employ the following quantity to describe the evolution of $\psiup$ at different temperature scales:
\begin{equation}
Y\coloneqq \frac{n_\psiup}{s}\propto
\frac{n_\psiup}{g_{*s}T^3}\,,
\end{equation}
where $s$ is the comoving entropy density $s\coloneqq S/V$,
\begin{equation}
s=\sum_i \frac{\rho_i+P_i-\mu_i n_i}{T_i}\simeq\frac{2\pi^2}{45}g_{*s}T^3
\quad\text{with}\quad g_{*s}
\coloneqq\sum_{\text{boson}} g_i^{(b)}
\left(\frac{T_{i}}{T}\right)^3
+\frac{7}{8}
\sum_\text{fermion} g_j^{(f)}
\left(\frac{T_{j}}{T}\right)^3\,.
\end{equation}
Here we have applied $P_i=\rho_i/3$ and $\mu_i\ll T_i$ in $s$ for relativistic matter, and $g_{*s}$ denotes
the entropic number of statistic degrees of freedom.
According to the continuity equation Eq.(\ref{Continuity Eq}) and the thermodynamic identities
\begin{equation}
\left.\frac{\partial P}{\partial T}\right|_\mu=s\,,\quad
\left.\frac{\partial P}{\partial \mu}\right|_T=n\,,
\end{equation}
one has
\begin{equation}
\frac{d(sa^3)}{dt}
=-\frac{\mu}{T}\frac{d(na^3)}{dt}\,,
\end{equation}
so the comoving entropy density $sa^3$ of
a particle species is conserved
when the comoving particle number density $n_\psiup a^3$ is conserved or the chemical potential $\mu$ is far smaller than the temperature. Thus, $d(sa^3)/dt=0=a^3(\dot s+3Hs)$,
$\dot s/s=-3H$, and the time derivative of $Y$ becomes
\begin{equation}\label{DM Boltzamann eq1}
\frac{dY}{dt}= \frac{\dot{n}_\psiup}{s}-\frac{\dot s}{s}Y= \frac{\dot{n}_\psiup}{s}+3HY=s^{-1}\left(\dot{n}_\psiup+3Hn_\psiup\right)\,.
\end{equation}
Substitute the simplified Boltzmann equation (\ref{simplified Boltzmann equation}) into Eq.(\ref{DM Boltzamann eq1}), and one obtains
\begin{equation}\label{DM Boltzamann eq3}
\frac{dY}{dt}= -s \left\langle\sigma v \right\rangle
\left(Y^2-Y^2_{\text{eq}} \right)\,.
\end{equation}
Now rewrite $dY/dt$ into $dY/dx$. Since
\begin{equation}\label{WIMP temperature vs time}
T=\left(\frac{30}{\pi^2 g_*}\rho\right)^{1/4}
=
\left\{
\frac{30\varepsilon^{2-2\beta}m_{\text{Pl}}^{2}
(-5\beta^2+8\beta-2 )}{ 32\pi^3 g_*(\beta-1 )}
\Big[3\beta(\beta-1 ) \Big]^\beta\right\}^{1/4}t^{-\beta/2} \;\;
\propto t^{-\beta/2}\,,
\end{equation}
thus ${\dot T}/{T}=-{\beta}/(2t)
=-H(t)=-H(x)=-H(m)x^{-2/\beta}$, and $\frac{dY}{dx}\frac{dx}{dT}\frac{dT}{dt}=
\frac{dY}{dx}\left(-x\right)
\frac{\dot T}{T}=
\frac{dY}{dx}\left(-x\right)\left(-H(m)x^{-2/\beta}\right)$, which recast Eq.(\ref{DM Boltzamann eq3}) into
\begin{equation}\label{DM Boltzamann eq4}
\frac{dY}{dx}=
-\frac{x^{\frac{2}{\beta} -1}}{H(m)}\left\langle\sigma v \right\rangle
s\left(Y^2-Y^2_{\text{eq}} \right)
=
-\frac{\left\langle\sigma v \right\rangle s}{Hx}
\left(Y^2-Y^2_{\text{eq}} \right)\,.
\end{equation}
Defining the annihilation rate of $\psiup$ as $\Gamma_\psiup\coloneqq n_{\text{eq}}
\left\langle\sigma v \right\rangle$, then Eq.(\ref{DM Boltzamann eq4}) can be rewritten into the form
\begin{equation}
\frac{x}{Y_{\text{eq}}}\frac{dY}{dx}=
-\frac{n_{\text{eq}}}{H}\left\langle\sigma v \right\rangle
\left[\left(\frac{Y}{Y_{\text{eq}}}\right)^2-1 \right]
=
-\frac{\Gamma_\psiup}{H}
\left[\left(\frac{Y}{Y_{\text{eq}}}\right)^2-1 \right]\,,
\end{equation}
which will be very useful in calculating the freeze-out temperature of cold relics in Sec.~\ref{Cold relic dark matter}.


\section{Hot/warm relic dark matter and light neutrinos}\label{Hot warm relic dark matter and light neutrinos}


\subsection{Generic bounds on $\psiup$ mass}

Having set up the modified cosmological dynamics and Boltzmann equations in $\mathscr{L}=\varepsilon^{2-2\beta}R^\beta+{16\pi}m_{\text{Pl}}^{-2}\mathscr{L}_m$ gravity, we will continue to investigate hot dark matter which is relativistic for the entire history of the Universe until now, and warm dark matter which is relativistic at the time of decoupling but become nonrelativistic nowadays.

In the relativistic regime $T\gg 3m_\psiup$ or equivalently $0<x\ll 3$, the abundance of $m_\psiup$ is given by
\begin{equation}\label{Y Equilibrium II}
\begin{split}
Y_{\text{eq}}=
Y_{\text{eq}}^{\text{(R)}}
&=\frac{45\zeta(3)}{2\pi^4}\frac{b_\psiup g_\psiup}{g_{*s}}
\simeq 0.2777 \frac{b_\psiup g_\psiup}{g_{*s}}\,,
\end{split}
\end{equation}
where $\zeta(3)=1.20206$, $b_\psiup=1$ for bosons and $b_\psiup=3/4$ for fermions.
$Y_{\text{eq}}$ only implicitly depends on $x$ through the evolution of $g_{*s}$ along the temperature scale. Then, the relic abundance is still given by $Y_{\text{eq}}$ at the time of freeze-out $x_f$:
\begin{equation}
Y_\infty\coloneqq Y(x\to\infty)
=Y_{\text{eq}}^{\text{(R)}}(x_f)=0.2777 \times \frac{b_\psiup g_\psiup}{g_{*s}(x_f)}\,.
\end{equation}
At the present time with $T_{\text{cmb}}=2.7255\, K$ \cite{Particle Data Group}, the entropy density is
\begin{equation}
s_0 = \frac{2\pi^2}{45}
g_{*s0}T^3_{\text{cmb}}=2891.2 \text{ cm}^{-3}\,,
\end{equation}
where in the minimal standard model with three generations of light neutrinos ($N_\nu=3$),
\begin{equation}
g_{*s0}=2+\frac{7}{8}\times2\times N_\nu\times
\left(\frac{T_{\nu0}}{T_{\text{cmb}}}\right)^3\simeq 3.9091\,.
\end{equation}
Thus, the present-day number density and
energy density of hot/warm relic $\psiup$ can be found by
\begin{equation}
n_{\psi0}=s_0 Y_\infty
=802.8862\times \frac{b_\psiup g_\psiup}{g_{*s}(x_f)} \text{ cm}^{-3} \,,
\end{equation}
\begin{equation}
\rho_{\psiup0}=m_\psiup n_{\psiup0}=
802.8862\times \frac{b_\psiup g_\psiup}{g_{*s}(x_f)}
\left(\frac{m_\psiup}{\text{eV}}\right)\; \frac{\text{eV}}{\text{cm}^3}\,,
\end{equation}
which, for $\rho_{\text{crit}}=1.05375\times 10^4 \,h^2 \text{ eV/cm}^3$,
correspond to the fractional energy density
\begin{equation}\label{Hot warm Omega h2}
\Omega_\psi h^2= \frac{\rho_{\psiup0}}{\rho_{\text{crit}}}h^2\times \frac{b_\psiup g_\psiup}{g_{*s}(x_f)} \left(\frac{m_\psiup}{\text{eV}}\right)
= 0.0762 \times \frac{b_\psiup g_\psiup}{g_{*s}(x_f)}
\left(\frac{m_\psiup}{\text{eV}}\right) \,.
\end{equation}
This actually stands for an attractive feature of
the paradigm of thermal relics: the current abundance $\Omega_\psi h^2$ of relic dark matter (hot, warm, or cold) can be predicted by $\psiup$'s microscopic properties like
mass, annihilation cross-section, and statistical degrees of freedom.

Since hot/warm relics can at most reach
the total dark matter density $\Omega_\psi h^2=0.1198\pm0.0026$ \cite{Particle Data Group}, $\Omega_\psi h^2$ has to satisfy
$\Omega_\psi h^2\lesssim0.1198$, and it follows from Eq.(\ref{Hot warm Omega h2}) that $m_\psiup$ is limited by the upper bound
\begin{equation}\label{mass bound upper}
m_\psiup\lesssim 1.5723\times \frac{g_{*s}(x_f)}
{b_\psiup g_\psiup}\text{ eV}\,.
\end{equation}
Moreover, particles of warm dark matter become nonrelativistic at present time, which imposes a lower bound to $m_\psiup$,
\begin{equation}\label{mass bound lower}
m_\psiup\gtrsim T_{\psiup0}=T_{\psiup f}\frac{a_f}{a_0}
=\left(\frac{g_{*s0}}{g_{*s}(x_f)} \right)^{1/3}
 T_{\text{cmb}}
=2.3496\times 10^{-4}\times \left(\frac{3.9091}{g_{*s}(x_f)} \right)^{1/3} \text{ eV} \,,
\end{equation}
where we have applied $g_{*s}^{1/3}aT=\text{constant}$ due to $sa^3=\text{constant}$.
Eqs.(\ref{mass bound upper}) and (\ref{mass bound lower}) lead to the mass bound for warm relics that
\begin{equation}\label{mass bound upper lower}
2.3496\times 10^{-4}\times \left(\frac{3.9091}{g_{*s}(x_f)} \right)^{1/3}
\lesssim \frac{m_\psiup}{\text{eV}} \lesssim 1.5723\times \frac{g_{*s}(x_f)} {b_\psiup g_\psiup}\,.
\end{equation}


\subsection{Example: light neutrinos as hot relics}

Light neutrinos are the most popular example of hot/warm dark matter \cite{Relics neutrinos I}. One needs to figure out the temperature $T_f^\nu$ and thus $g_{*s}(T=T_f^\nu)$ when neutrinos freeze out from the cosmic plasma.
The decoupling occurs when the Hubble expansion rate $H$ balances neutrinos' interaction rate $\Gamma_{\nu}$. For the cosmic expansion, it is convenient to write Eq.(\ref{Hubble evolution}) into
\begin{equation}\label{DM Hubble evolution II}
\begin{split}
H= 0.2887\times\sqrt{\frac{\beta}{\beta-1}}\times\left(\sqrt{\frac{(\beta-1)\,g_*}
{-5\beta^2+8\beta-2}}\right)^{1/\beta}
\left(0.7164\cdot T^2_{\text{MeV}}\right)^{1/\beta}
 \varepsilon_{s}^{1 -\frac{1}{\beta}}\text{ [1/s]}\,,
\end{split}
\end{equation}
where $T_{\text{MeV}}$ refers to the value of temperature in the unit of MeV, $T=T_{\text{MeV}}\times\text{ [1\,MeV]}$, $\varepsilon_{s}$ is the value of $\varepsilon$ in the unit of [1/s], and
numerically ${T^2}/{m_{\text{Pl}}}={T^2_{\text{MeV}}}/{8.0276}$ [1/s].

On the other hand, the event of neutrino decoupling
actually indicates the beginning of primordial nucleosynthesis, when neutrinos are in chemical
and kinetic equilibrium with photons, nucleons and electrons via weak
interactions  and elastic scattering. The interaction rate $\Gamma_{\nu}$ is \cite{DM Kolb Turner Book}
\begin{equation}\label{DM Neutrino interaction rate}
\Gamma_{\nu}
\simeq 1.3 G_F^2 T^5 \simeq 0.2688\,
T^5_{\text{MeV}} \text{ [1/s]}\,,
\end{equation}
where $G_F$ is Fermi's constant in beta decay
and generic weak interactions, and
$G_F=1.1664\times 10^{-11} \text{MeV}^{-2}$.  Neutrinos decouple when $\Gamma_{\nu}=H$,
and according to Eqs.(\ref{DM Hubble evolution II}) and (\ref{DM Neutrino interaction rate}),
the weak freeze-out temperature $T_\nu^f$ is the solution to
\begin{equation}\label{DM Neutrino decoupling temperature}
T^{5-2/\beta}_{\text{MeV}} = 1.0741\,\times
\sqrt{\frac{\beta}{\beta-1}}\times\left(0.7164\cdot
\sqrt{\frac{(\beta-1)\,g_*}
{-5\beta^2+8\beta-2}}\right)^{1/\beta}
\varepsilon_{s}^{1 -1/\beta}\,.
\end{equation}
Figs.~\ref{DMFigLightNeutinoI} and \ref{DMFigLightNeutinoII} have shown the dependence of $T_\nu^f$ on $\beta$ for $\varepsilon=1 \text{ sec}^{-1}=6.58\times 10^{-22}$ MeV and $\varepsilon=m_{\text{Pl}}=1.2209\times 10^{-22}$ MeV, respectively. Fig.~\ref{DMFigLightNeutinoII} clearly illustrates that $T_\nu^f$ spreads from 1.3030 MeV to over 1000 MeV, which goes far beyond the scope of $1\sim 10$ MeV; thus, as shown in Table ~\ref{Table 1 gs}, $g_{*s}$ varies and the mass bound $\Sigma m_\nu$ in light of Eq.(\ref{mass bound upper lower}) is both $\beta-$dependent and $\varepsilon-$dependent.

\vspace{-5mm}

\begin{tabular}{c}

\begin{minipage}{\linewidth}
 \makebox[1\linewidth]{ \includegraphics[keepaspectratio=true,scale=0.5]{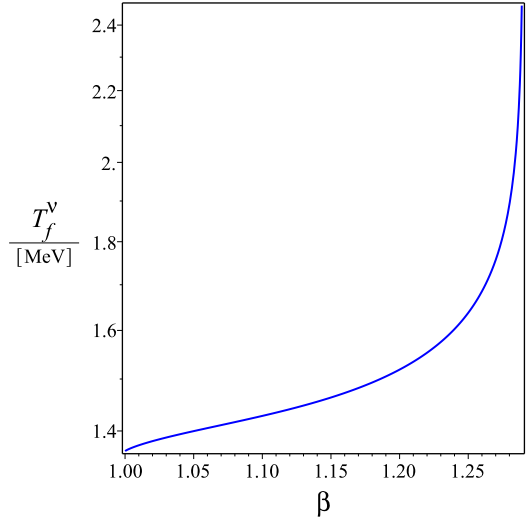}}
 \vspace{-6.5cm}
\captionof{figure}
{$T_\nu^f$ (in MeV)
for $\varepsilon=1 \text{ sec}^{-1}=6.58\times 10^{-22}$ MeV}\label{DMFigLightNeutinoI}
\end{minipage}
\\

\begin{minipage}{\linewidth}
 \hspace{-10mm}\makebox[1\linewidth]{ \includegraphics[keepaspectratio=true,scale=0.48]{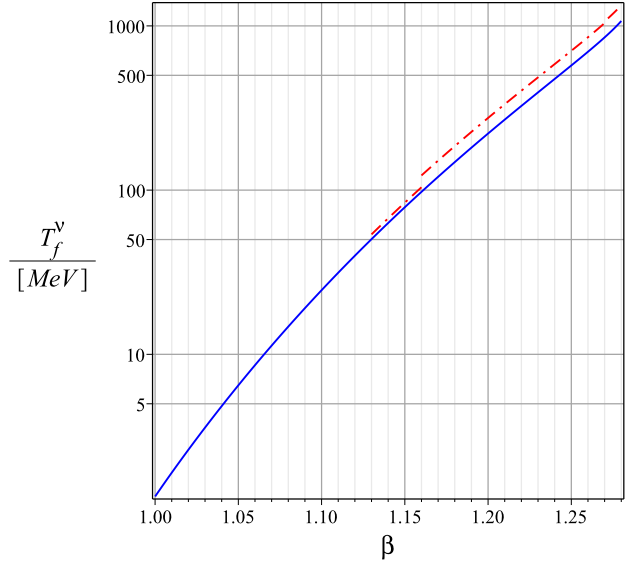}}
\vspace{-5cm}
\captionof{figure}{$T_\nu^f$ (in MeV) for
$\varepsilon=m_{\text{Pl}}=1.2209\times 10^{-22}$ MeV}\label{DMFigLightNeutinoII}
\end{minipage}

\end{tabular}\\

\vspace{5mm}

\begin{table}
\begin{center}
\renewcommand\arraystretch{1.69}
\caption{$g_{*s}$ for the $T_f^\nu$ in Fig.~\ref{DMFigLightNeutinoII}, based on the data of Particle Data Group. Note that between 100$\sim$200 MeV, $g_{*s}$ is also subject to the phase transition of quantum chromodynamics for strange quarks.}\label{Table 1 gs}
\begin{tabular}{l|l|l}
  \hline
  Temperature & Temperature (in MeV) & $g_{*s}$  \\ \hline
  $m_e<T<m_s \text{ (strange)}$ & $0.5110<T<95$ & 43/4  \\
  $m_s <T<m_\mu $ & $95<T<105.6584 $ & 57/4 \\
  $m_\mu <T<m_\pi$ &  $105.6584 <T< 134.9766$ &  69/4\\
  $m_\pi<T<T_c$ &  $134.9766 <T<T_c $ &  205/4\\
  $T_c<T<m_c \text{ (charm)}$ &  $ T_c<T<1275$ &  247/4\\
  $m_c<T<m_\tau $ &  $1275<T<1776.82 $ &  289/4\\
  $m_\tau <T<m_b \text{ (bottom)}$ &  $1776.82 <T<4180$ &  303/4\\

  \hline
\end{tabular}
\end{center}
\end{table}

\section{Cold relic dark matter}\label{Cold relic dark matter}

Now let's consider cold dark matter which is already
nonrelativistic at the time of decoupling.
In the nonrelativistic regime $T\ll 3m_\psiup$ or
equivalently $x\gg3$, the number density and entropy density are given by
\begin{equation}
n_\psiup =g_\psiup\left(\frac{m_\psiup^2}{2\pi}\right)^{3/2} x^{-3/2} e^{-x}\,,\quad
s=\frac{2\pi^2}{45}g_{*s}m_\psiup^3 x^{-3}
=s(m)\, x^{-3}\,,
\end{equation}
so one obtains the equilibrium abundance of nonrelativistic $\psiup$ particles
\begin{equation}\label{Y Equilibrium I}
\begin{split}
Y_{\text{eq}}=Y_{\text{eq}}^{\text{(NR)}}
=\frac{45}{4\pi^4}\left(\frac{\pi}{2}\right)^{1/2}
\frac{g_\psiup}{g_{*s}}x^{3/2}e^{-x}
\simeq 0.1447\times\frac{g_\psiup}{g_{*s}}x^{3/2}e^{-x}\,.
\end{split}
\end{equation}
Thus, $n_\psiup$ and $Y_{\text{eq}}=Y_{\text{eq}}^{\text{(NR)}}$ are exponentially
suppressed when the temperature drops below $m_\psiup$.
Moreover, since cold relics are nonrelativistic when freezing out,
one can expand the thermally averaged cross-section by $\left\langle\sigma v \right\rangle = c_0
+c_1 v^2+c_2 v^4+\cdots+c_qv^{2q}+\cdots$, where $c_0$ corresponds to the decay
channel of $s-$wave, $c_1$ to $p-$wave, $c_2$ to $d-$wave,
and so forth; recalling that $\left\langle\sigma v \right\rangle\sim \sqrt{T}$
in light of the Boltzmann velocity distribution, thus
the annihilation cross-section can be expanded by the variable $x$ into
\begin{equation}
\left\langle\sigma v \right\rangle
=\left\langle\sigma v \right\rangle_0 x^{-n}
\quad \text{with}\quad n=q/2\,.
\end{equation}
Then the Boltzmann equation (\ref{DM Boltzamann eq4}) becomes
\begin{equation}\label{Integrate Y I}
\frac{dY}{dx}=
-\frac{s(m)\left\langle\sigma v \right\rangle_0}
{H(m)} x^{\frac{2}{\beta} -4-n} \left(Y^2 - Y^2_{\text{eq}}\right)=
-\frac{s(m)\left\langle\sigma v \right\rangle_0}{H(m)} x^{\frac{2}{\beta} -4-n}
 \left[Y^2-0.0209\left(\frac{g_\psiup}{g_{*s}}\right)^2x^3 e^{-2x} \right]\,,
\end{equation}
where
\begin{equation}
\begin{split}
\frac{s(m)\left\langle\sigma v \right\rangle_0}{H(m)}  = \frac{1.519525}{\left(5.750944 \right)^{1/\beta}}
\sqrt{\frac{\beta-1}{\beta}}
\left(\!\sqrt{\frac{-5\beta^2+8\beta-2}{\beta-1}}\right)^{1/\beta}
\frac{g_{*s}}{(\!\sqrt{g_*})^{1/\beta}}\,\varepsilon^{\frac{1}{\beta}-1} \, m^{3-\frac{2}{\beta}}\, m_{\text{Pl}}^{1/\beta}\left\langle\sigma v \right\rangle_0\,.
\end{split}
\end{equation}

Though initially in equilibrium $Y\approx Y_{\text{eq}}=Y_{\text{eq}}^{\text{(NR)}}$, the actual abundance $Y$ gradually
departures from the equilibrium value $Y_{\text{eq}}^{\text{(NR)}}$
as the temperature decreases; $Y$ freezes out and escapes the exponential
Boltzmann suppression when the interaction rate $\Gamma_\psiup$
equates the cosmic expansion rate $H$.
Transforming Eq.(\ref{Integrate Y I}) into the form
\begin{equation}\label{Integrate Y III}
\frac{x}{Y_{\text{eq}}}\frac{dY}{dx}=
-\frac{s(m)\left\langle\sigma v \right\rangle_0}{H(m)}
Y_{\text{eq}} \left[\left(\frac{Y}{Y_{\text{eq}}}\right)^2- 1\right] x^{\frac{2}{\beta}-3-n}
=
-\frac{\Gamma_\psiup}{H}
\left[\left(\frac{Y}{Y_{\text{eq}}}\right)^2-1 \right]\,,
\end{equation}
and the coupling condition $\Gamma_\psiup(x_f)=H(x_f)$
at the freeze-out temperature $T_f^\psiup=m_\psiup/x_f$ yields
\begin{equation}
\frac{\Gamma_\psiup}{H}(x_f)=1 = \frac{s(m)\left\langle\sigma v \right\rangle_0}{H(m)}
  Y_{\text{eq}}  x^{\frac{2}{\beta}-3-n}\simeq 0.1447 \frac{s(m)\left\langle\sigma v \right\rangle_0}{H(m)}
 \frac{g_\psiup}{g_{*s}} x^{\frac{2}{\beta}-3/2-n} e^{-x} \,.
\end{equation}
Thus, it follows that
\begin{equation}\label{CDM xf Eq}
\begin{split}
e^{x_f}&=0.1447 \frac{s(m)\left\langle\sigma v \right\rangle_0}{H(m)}
\frac{g_\psiup}{g_{*s}} x^{\frac{2}{\beta}-3/2-n}\\
&=\frac{0.2199}{\left(5.7509 \right)^{1/\beta}}
\sqrt{\frac{\beta-1}{\beta}}
\left(\!\sqrt{\frac{-5\beta^2+8\beta-2}{\beta-1}}\right)^{1/\beta}
\frac{g_\psiup}{(\!\sqrt{g_*})^{1/\beta}}\,\varepsilon^{\frac{1}{\beta}-1}
\, m^{3-\frac{2}{\beta}}\, m_{\text{Pl}}^{1/\beta}\left\langle\sigma v \right\rangle_0
x_f^{\frac{2}{\beta}-3/2-n}\,.
\end{split}
\end{equation}
After taking the logarithm of both side,
Eq.(\ref{CDM xf Eq}) can be iteratively solved to obtain
\begin{align}\label{Cold relics xf}
x_f = &\ln\left[\frac{0.2199}{\left(5.7509 \right)^{1/\beta}}
\sqrt{\frac{\beta-1}{\beta}}
\left(\!\sqrt{\frac{-5\beta^2+8\beta-2}{\beta-1}}\right)^{1/\beta}
\frac{g_\psiup}{(\!\sqrt{g_*})^{1/\beta}}\,\varepsilon^{\frac{1}{\beta}-1} \,
m^{3-\frac{2}{\beta}}\, m_{\text{Pl}}^{1/\beta}\left\langle\sigma v \right\rangle_0 \right]\\
& +\left(\frac{2}{\beta}-\frac{3}{2}-n\right)\ln \left[\ln\Bigg(\frac{0.2199}{\left(5.7509 \right)^{1/\beta}}
\sqrt{\frac{\beta-1}{\beta}}
\left(\!\sqrt{\frac{-5\beta^2+8\beta-2}{\beta-1}}\right)^{1/\beta}
\frac{g_\psiup}{(\!\sqrt{g_*})^{1/\beta}}\,\varepsilon^{\frac{1}{\beta}-1} \,
m^{3-\frac{2}{\beta}}\, m_{\text{Pl}}^{1/\beta}\left\langle\sigma v \right\rangle_0 \Bigg)  \right]\nonumber\\
& +\left(\frac{2}{\beta}-\frac{3}{2}-n\right)\ln \Bigg[\cdots \cdots \Bigg]\,,\nonumber
\end{align}
where $g_*$ has been treated as a constant, as the time scale over which $g_*$ evolves is much greater than the time interval near $x_f$.


\subsection{Abundance $Y$ before freeze-out}
\vspace{2mm}

To work out the actual abundance $Y$ before the decoupling of $\psi$, employ a new quantity $\Delta\coloneqq Y-Y_{\text{eq}}$, and then Eq.(\ref{Integrate Y I}) can be recast into
\begin{equation}\label{Integrate Y II-1}
\frac{d\Delta}{dx}=
-\frac{s(m)\left\langle\sigma v \right\rangle_0}{H(m)} x^{\frac{2}{\beta} -4-n}
\Delta\left(\Delta+2Y_{\text{eq}} \right)-\frac{dY_{\text{eq}}}{dx}\,.
\end{equation}
In the high-temperature regime $x\ll x_f$ before $\psi$ freezes out,
$Y$ is very close to $Y_{\text{eq}}$, so that $\Delta\ll Y_{\text{eq}}$ and $d\Delta/dx\ll-dY_{\text{eq}}/dx$.
With $Y_{\text{eq}}=Y_{\text{eq}}^{\text{(NR)}}$ in Eq.(\ref{Y Equilibrium I}),  Eq.(\ref{Integrate Y II-1}) can be algebraically solved to obtain
\begin{equation}
\begin{split}
\Delta&=-\frac{dY_{\text{eq}}}{dx}
\frac{H(m)}{s(m)\left\langle\sigma v \right\rangle_0}
\frac{x^{n+4-\frac{2}{\beta}}}{2Y_{\text{eq}}+\Delta} 
=\left(1-\frac{3}{2x}\right)
\frac{H(m)}{s(m)\left\langle\sigma v \right\rangle_0}
\frac{x^{n+4-\frac{2}{\beta}}}{2+Y_{\text{eq}}/\Delta} 
\simeq \left(1-\frac{3}{2x}\right)
\frac{H(m)}{2s(m)\left\langle\sigma v \right\rangle_0}
x^{n+4-\frac{2}{\beta}},
\end{split}
\end{equation}
and consequently
\begin{equation}\label{CDM Y before xf}
\begin{split}
Y=\Delta+Y_{\text{eq}}
&=\left(1-\frac{3}{2x}\right)
\frac{H(m)}{s(m)\left\langle\sigma v \right\rangle_0}
\frac{x^{n+4-\frac{2}{\beta}}}{2+Y_{\text{eq}}/\Delta}
+\frac{45}{4\pi^4}\left(\frac{\pi}{2}\right)^{1/2}
\frac{g_\psiup}{g_{*s}}x^{3/2}e^{-x}\\
&\simeq  \left(1-\frac{3}{2x}\right)
\frac{H(m)}{2s(m)\left\langle\sigma v \right\rangle_0}
x^{n+4-\frac{2}{\beta}}
+ 0.1447\times\frac{g_\psiup}{g_{*s}}x^{3/2}e^{-x}\,.
\end{split}
\end{equation}


\subsection{Freeze-out abundance $Y_\infty$}
\vspace{2mm}

After the decoupling of $\psi$ particles, the actual number density $n_\psiup$ becomes
much bigger than the ideal equilibrium value $n_\psiup^{\text{eq}}$. One has
$Y \gg Y_{\text{eq}}$, $Y \approx \Delta$, and the differential equations (\ref{Integrate Y I}) or (\ref{Integrate Y II-1}) leads to
\begin{equation}\label{Integrate Y IV}
\frac{dY}{dx}=
-\frac{s(m)\left\langle\sigma v \right\rangle_0}{H(m)} x^{\frac{2}{\beta} -4-n}\,Y^2
\quad\text{or}\quad
\frac{dY}{dx}=
-\frac{s(m)\left\langle\sigma v \right\rangle_0}{H(m)}
x^{\frac{2}{\beta} -4-n}\,Y^2\,,
\end{equation}
which integrates to yield the freeze-out abundance $Y_\infty\coloneqq Y(x=x_f)\approx Y(x\to\infty)$ that
\begin{equation}\label{CDM Y after xf}
\begin{split}
Y_\infty&=\left(3+n-\frac{2}{\beta}\right) \frac{H(m)}{s(m)\left\langle\sigma v \right\rangle_0}
x_f^{3+n-\frac{2}{\beta}}\\
&=\frac{\left(3+n-\frac{2}{\beta}\right)x_f^{3+n-\frac{2}{\beta}}}
{\frac{1.5195}{\left(5.7509 \right)^{1/\beta}}
\sqrt{\frac{\beta-1}{\beta}}
\left(\!\sqrt{\frac{-5\beta^2+8\beta-2}{\beta-1}}\right)^{1/\beta}
\frac{g_{*s}}{(\!\sqrt{g_*})^{1/\beta}}\,\varepsilon^{\frac{1}{\beta}-1} \, m^{3-\frac{2}{\beta}}\, m_{\text{Pl}}^{1/\beta}\left\langle\sigma v \right\rangle_0  }\,.
\end{split} 
\end{equation}
Following $Y_\infty$, the number density and energy density of $\psi$
are directly are directly found to be
\begin{equation}
n_{\psiup0}=s_0 Y_\infty =\frac{2891.2\left(3+n-\frac{2}{\beta}\right)
x_f^{3+n-\frac{2}{\beta}}}
{\frac{1.5195}{\left(5.7509 \right)^{1/\beta}}
\sqrt{\frac{\beta-1}{\beta}}
\left(\!\sqrt{\frac{-5\beta^2+8\beta-2}{\beta-1}}\right)^{1/\beta}
\frac{g_{*s}}{(\!\sqrt{g_*})^{1/\beta}}\,\varepsilon^{\frac{1}{\beta}-1} \,
m^{3-\frac{2}{\beta}}\, m_{\text{Pl}}^{1/\beta}\left\langle\sigma v \right\rangle_0  }
\text{ cm}^{-3}\,,
\end{equation}
\begin{equation}
\rho_{\psiup0}=m_\psiup n_{\psiup0}
=\frac{2891.2\left(3+n-\frac{2}{\beta}\right)x_f^{3+n-\frac{2}{\beta}}}
{\frac{1.5195}{\left(5.7509 \right)^{1/\beta}}
\sqrt{\frac{\beta-1}{\beta}}
\left(\!\sqrt{\frac{-5\beta^2+8\beta-2}{\beta-1}}\right)^{1/\beta}
\frac{g_{*s}}{(\!\sqrt{g_*})^{1/\beta}}\,\varepsilon^{\frac{1}{\beta}-1} \,
m^{2-\frac{2}{\beta}}\, m_{\text{Pl}}^{1/\beta}\left\langle\sigma v \right\rangle_0  }
\; \frac{\text{eV}}{\text{cm}^3}\,,
\end{equation}
which gives rise to the fractional energy density
\begin{equation}\label{CDM Omegah2}
\Omega_\psi h^2= \frac{\rho_{\psiup0}}{\rho_{\text{crit}}}h^2
=\frac{2743.7248 \left(3+n-\frac{2}{\beta}\right)x_f^{3+n-\frac{2}{\beta}}}
{\frac{1.5195}{\left(5.7509 \right)^{1/\beta}}
\sqrt{\frac{\beta-1}{\beta}}
\left(\!\sqrt{\frac{-5\beta^2+8\beta-2}{\beta-1}}\right)^{1/\beta}
\frac{g_{*s}}{(\!\sqrt{g_*})^{1/\beta}}\,\varepsilon^{\frac{1}{\beta}-1} \,
m^{2-\frac{2}{\beta}}\, m_{\text{Pl}}^{1/\beta}\left\langle\sigma v \right\rangle_0  } \,.
\end{equation}


\noindent Unlike Eq.(\ref{Hot warm Omega h2}) for hot/warm relics,
the relic density $\Omega_\psi h^2$ for cold dark matter is not only much more sensitive to the
temperature of cosmic plasma, but also relies on the annihilation cross-section.


\section{Example: Fourth generation massive neutrinos and Lee-Weinberg bound}
\label{Example Fourth generation massive neutrinos and Lee-Weinberg bound}

An example of cold relics can be the hypothetical fourth generation
massive neutrinos  \cite{Lee Weinberg bound, Lee Weinberg bound II, Relics neutrinos I}. For the Dirac-type
neutrinos whose annihilations are dominated by $s-$wave $(n = 0)$, the interaction cross-section reads
\begin{equation}
\left\langle\sigma v \right\rangle_0\simeq G_F^2 m^2
= 1.3604\times 10^{-10}
\left(\frac{m_\psiup}{\text{GeV}}\right)^2 \text{ GeV}^{-2}
\end{equation}
where $G_F$ is Fermi's constant in beta decay and generic
weak interactions, and $G_F=1.16637\times 10^{-5} \text{GeV}^{-2}$.  Then with $g_\psiup=2$ and $g_*\sim 60$,
the neutrinos decouple at
\begin{align}\label{Lee Weinberg Cold relics xf}
\tilde{x}_f = &\ln\left[\frac{0.5983\times 10^{-10} }{\left(44.5463 \right)^{1/\beta}}
\sqrt{\frac{\beta-1}{\beta}}
\left(\!\sqrt{\frac{-5\beta^2+8\beta-2}{\beta-1}}\right)^{1/\beta}
 \,\varepsilon^{\frac{1}{\beta}-1} \, m^{5-\frac{2}{\beta}}\, m_{\text{Pl}}^{1/\beta} \right]\\
& +\left(\frac{2}{\beta}-\frac{3}{2}-n\right)\ln \left[\ln\Bigg(\frac{0.5983\times 10^{-10} }{\left(44.5463 \right)^{1/\beta}}
\sqrt{\frac{\beta-1}{\beta}}
\left(\!\sqrt{\frac{-5\beta^2+8\beta-2}{\beta-1}}\right)^{1/\beta}
 \,\varepsilon^{\frac{1}{\beta}-1} \, m^{5-\frac{2}{\beta}}\, m_{\text{Pl}}^{1/\beta} \Bigg)  \right]\nonumber\\
& +\left(\frac{2}{\beta}-\frac{3}{2}-n\right)\ln \Bigg[\cdots \cdots \Bigg]\,,\nonumber
\end{align}
which, through Eq.(\ref{CDM Omegah2}), gives rise to the fractional energy density
\begin{equation}\label{CDM Omegah2}
\Omega_\psi h^2= \frac{\rho_{\psiup0}}{\rho_{\text{crit}}}h^2
=\frac{2743.7248\times 10^{10}\times \left(3+n-\frac{2}{\beta}\right)\tilde{x}_f^{3+n-\frac{2}{\beta}}}
{\frac{0.5983}{\left(44.5463 \right)^{1/\beta}}
\sqrt{\frac{\beta-1}{\beta}}
\left(\!\sqrt{\frac{-5\beta^2+8\beta-2}{\beta-1}}\right)^{1/\beta}
 \,\varepsilon^{\frac{1}{\beta}-1} \, m^{5-\frac{2}{\beta}}\, m_{\text{Pl}}^{1/\beta} } \,.
\end{equation}%
With the same amount of anti-particles, we finally have
$\Omega_{\psiup\bar\psiup0} h^2= 2\Omega_\psi h^2\lesssim 0.1198$. Thus the Lee-Weinberg
bound \cite{Lee Weinberg bound, Lee Weinberg bound II} for massive neutrinos are relaxed in $\mathscr{L}=\varepsilon^{2-2\beta}R^\beta+{16\pi}m_{\text{Pl}}^{-2}\mathscr{L}_m$ gravity.

\section{Conclusions}

In this paper, we have comprehensively investigated the thermal
relics as hot, warm and cold dark matter in  $\mathscr{L}=\varepsilon^{2-2\beta}R^\beta+{16\pi}m_{\text{Pl}}^{-2}\mathscr{L}_m$ gravity. When light neutrinos act as hot and warm neutrinos, the upper limit of neutrino mass $\Sigma m_\nu$ relies on the value of $\beta$ and the choice of $\varepsilon$. For cold relics, we have derived the freeze-out temperature $T_f=m/x_f$ in
Eq.(\ref{Cold relics xf}), $Y$ before the freeze-out in Eq.(\ref{CDM Y before xf}), the freeze-out value $Y_\infty$ in Eq.(\ref{CDM Y after xf}), and the dark-matter fractional density $\Omega_\psi h^2$ in Eq.(\ref{CDM Omegah2}).
Note that we focused on power-law $f(R)$ gravity because unlike the approximated power-law ansatz $a=a_0 t^\alpha$ ($\alpha=\text{constant}>0$) for generic $f(R)$ gravity, $a=a_0 t^{\beta/2}$ is an exact solution to
$\mathscr{L}=\varepsilon^{2-2\beta}R^\beta+{16\pi}m_{\text{Pl}}^{-2}\mathscr{L}_m$ gravity for the radiation-dominated Universe; for GR with $\beta\to 1^+$, Eq.(\ref{Exact solution}) reduces to recover the behavior $a\propto  t^{1/2}$ which respects $3\dot{a}^2/a^2=-{8\pi}m_{\text{Pl}}^{-2} \rho_0 a^{-4}$.

When light neutrinos serve as hot/warm relics, the entropic number
of statistical degrees of freedom $g_{*s}$ at freeze-out and thus the predicted fractional energy density $\Omega_\psiup h^2$ are $\beta-$dependent, which
relaxes the standard mass bound $\Sigma m_\nu$.  For cold relics, by exactly solve the simplified Boltzmann equation in both relativistic and nonrelativistic regimes, we show that the Lee-Weinberg bound for the mass of heavy
neutrinos can be considerably relaxed, and the ``WIMP miracle'' for weakly interacting massive particles (WIMPs) gradually becomes invalid when $\beta$ departs $\beta=1^+$.    The whole framweork reduces to become that of GR in the limit $\beta\to 1^+$.


\end{document}